# Экстремумы производства энтропии и второй закон в релеевском газе.

Таджибаев П.А.,          Таджибаев Д.П.

Национальный университет Узбекистана имени Мирзо Улугбека.

learnlink@rambler.ru

Из полученных ранее решений уравнения Фоккера - Планка для релеевского газа (малой примеси тяжелых частиц в термостате легких частиц) с источниками и без них, вычислены производства энтропии. Показано, что в системе без источника (изолированная система) выполняется теорема Пригожина и в системе с источниками (открытая система) реализация принципа Циглера (МЕРР) зависит от направления релаксации. В открытой системе производства энтропии компенсируют отрицательным производства энтропии, т.е. производством негэнтропии. Алгебраическая сумма производств энтропии и негэнтропии вводится как обобщенное производство энтропии. Из баланса производств энтропии и негэнтропии в открытой системе сформулирован возможный вариант второго закона для открытых систем в форме: «При релаксации открытой системы в неравновесное стационарное состояние, обобщенное производство энтропии уменьшается по абсолютной величине и равно нулю в неравновесном стационарном состоянии".

Ключевые слова: уравнение Фоккера-Планка, теорема Пригожина, принцип максимума производства энтропии, второй закон, негэнтропия.

## Введение.

Создание феноменологической неравновесной термодинамики в линейном варианте в настоящее время завершена. Следующей задачей в развитие термодинамики является создание нелинейной неравновесной термодинамики [1]. В линейной неравновесной термодинамики была сформулирована Пригожиным И.Р. теорема о минимуме производства энтропии [2], которая в литературе часто цитируют как принцип минимального производства энтропии - ПминПЭ. Также как в своё время из принципа наименьшего действия был развит лагранжев формализм классической механики, можно было надеется, что на основе ПМинПЭ удастся создать нелинейную неравновесную термодинамику. Однако ПМинПЭ, как оказалось, применим в очень узкой области и неприменим в нелинейной области, это отмечал и сам Пригожин И.Р.[2]. В 1963 году Г. Циглером [3] был сформулирован принцип максимума производства энтропии – ПМаксПЭ, который имеет более широкую область применения. Однако, оба принципа являются феноменологическими и основаны на представление производства энтропии в виде произведения термодинамических сил и потоков. Поэтому эти подходы ограничены в своих возможностях, и более того эти принципы не получили пока адекватного статистического и кинетического обоснования. Наиболее широкие и глубокие исследования в неравновесной термодинамике позволяют провести кинетические подходы. Однако трудности решения соответствующих кинетических уравнений ограничивают применение кинетических методов в неравновесной термодинамике. Следует отметить, что в [4] уравнение Больцмана решено в приближении Бхатнагара – Гросса - Крука (модель сильных столкновений) и доказано ПМаксПЭ. Для более подробного анализа ПМаксПЭ модель сильных столкновений является грубым приближением. О роли ПМаксПЭ в необратимом процессе очень хорош сказано в [5] «…принцип максимального производства энтропии…" заставляет" систему выбирать не только наиболее вероятное из своих возможных микросостояний, но и наиболее вероятную траекторию движения к этому микросостоянию».

Целью настоящей работы является исследование экстремумов производства энтропии и свойств производства энтропии в нелинейной неравновесной термодинамике. Данное исследование проводится для модельной задачи, кинетическое уравнение которой имеет аналитическое решение.

1. Теорема Пригожина или ПМинПЭ.

Рассмотрим релеевский газ (малая примесь тяжелых частиц в термостате из легких частиц). Из-за малой концентрации тяжёлой компоненты и большой разности масс частиц смеси, установление равновесия в тяжелой компоненте является медленным процессом и происходит на фоне максвелловского распределения частиц термостата. В [6] интеграл упругих столкновение жестких сфер в уравнение Л. Больцмана разложением по степеням $(m_L/m)^{1/2}$, где $m_L$ и $m$ массы частиц термостата и примеси, соответственно, приведено к виду дифференциального оператора Фоккера – Планка вида

$$\frac{\partial F}{\partial \tau} = \frac{\partial}{\partial x}\left[\left(x - \frac{1}{2}\right)F + x\frac{\partial F}{\partial x}\right] \qquad (1)$$

с начальным и граничными условиями
$$F(x,0) = \varphi(x) \text{ и } j = x\frac{\partial F}{\partial x} + \left(x - \frac{1}{2}\right)F = 0, \text{ при } x = 0, x \to \infty, \qquad (2)$$

где $F(x,\tau)$ – функция распределения (ФР) тяжелых частиц, $x = \varepsilon/kT_L$, $\tau = t/\tau_R$ - безразмерные энергии и время, $\varepsilon$ – энергия тяжелых частиц, $T_L$ - температура термостата, $\tau_R$ - время релаксации релеевского газа [6]. Решение (1) с начальным и граничными условиями (2) имеет вид [6]

$$F(x,\tau) = \frac{2n_0}{\sqrt{\pi}} x^{1/2} e^{-x} + x^{1/2} e^{-x} \int_0^\infty \varphi(\xi) \sum_{m=1}^\infty \frac{\Gamma(m+1)}{\Gamma(m+3/2)} L_m^{1/2}(x) L_m^{1/2}(\xi) e^{-m\tau} d\xi , \quad (3)$$

где $L_m^{1/2}(x)$ – полиномы Лагерра, $\Gamma(m)$ – гамма функция, $n_0$ – начальное число частиц. Решение (3) применимо практически при всех x за исключением $(m/m_L)^{1/2} < x < (m_L/m)^{1/2}$, где уравнение Фоккера – Планка не применимо. Определив энтропию по Больцману в виде

$$S(\tau) = -k \int_0^\infty \left( \ln \frac{F(x,\tau)}{\sqrt{x}} - 1 \right) F(x,\tau) dx , \quad (4)$$

где k – постоянная Больцмана, для изменения энтропии получим

$$\frac{dS}{d\tau} = -k \int_0^\infty \left( \ln \frac{F(x,\tau)}{\sqrt{x}} \right) \frac{\partial F}{\partial \tau} dx \quad (5)$$

Подставив (3) в (5) имеем

$$\frac{dS}{d\tau} = k n_0 L_1^{1/2}(x) e^{-\tau} + k \frac{\sqrt{\pi}}{2} \sum_{m=1}^\infty \frac{m\Gamma(m+1)}{\Gamma(m+3/2)} \left[ \int_0^\infty \varphi(\xi) L_m^{1/2}(\xi) d\xi \right]^2 e^{-2m\tau} , \quad (6)$$

Из (6) следует, что

$$\sigma(\tau) = k \frac{\sqrt{\pi}}{2} \sum_{m=1}^\infty \frac{m\Gamma(m+1)}{\Gamma(m+3/2)} \left[ \int_0^\infty \varphi(\xi) L_m^{1/2}(\xi) d\xi \right]^2 e^{-2m\tau} \geq 0. \quad (7)$$

Как видно из (7), производство энтропии, в соответствие со вторым законом и теоремой Пригожина, в процессе релаксации из неравновесного в равновесное состояние положительна и убывает, достигая минимального нулевого значения в равновесии.

2. Принцип Циглера или ПМаксПЭ.

Для анализа принципа Циглера рассмотрим релаксацию начальной равновесной функции распределения релеевских частиц в конечное неравновесное стационарное распределение под действием внешних источников частиц. В качестве, внешнего источника выберем $\delta$ – образный источник тяжелых частиц. Этот источник вводит в систему моноэнергетические тяжёлые частицы с энергией $x_0$, которые, затем сталкиваясь с частицами термостата формируют текущую функцию распределения. Чтобы в системе не накапливались частицы введем «химическую» реакцию для вывода из системы частиц с текущей ФР.

Уравнение Фоккера – Планка в этом случае имеет вид

$$\frac{\partial F}{\partial \tau} = \frac{\partial}{\partial x} \left[ \left( x - \frac{1}{2} \right) F + x \frac{\partial F}{\partial x} \right] - K \tau_R F(x,\tau) + \eta \tau_R \delta(x - x_0), \quad (8)$$

с начальным и граничными условиями (2), где $x_0 = \varepsilon_0 / kT_L$ – безразмерная энергия частиц $\delta$ – источника, $\varphi(x) = \frac{2n_0}{\sqrt{\pi}} x^{1/2} e^{-x}$ - начальная равновесная функция распределения, $\varepsilon_0$ - энергии частиц $\delta$ – источника, k – постоянная Больцмана, $n_0$ - начальное число частиц, $K$ – константа химической реакции и $\eta$ – мощность $\delta$ – источника. Для данной модели аналитическое решение, в виде разложения по полиномам Лагерра, было получено ранее [7]. При условии баланса числа частиц $n_0 = \eta / K$, с начальной равновесной ФР, это решение имеет вид

$$F(x,\tau) = \frac{2n_0}{\sqrt{\pi}} x^{1/2} e^{-x} + \eta \tau_R K \tau_R x^{1/2} e^{-x} \sum_{m=1}^{\infty} \frac{\Gamma(m)}{\Gamma(m+3/2)} L_m^{1/2}(x) L_m^{1/2}(x_0) \left(1 - e^{-(m+K\tau_R)\tau}\right), \quad (9)$$

Определив энтропию по Больцману в виде (4) для изменения энтропии получим (5). Подставив (8) в (5) имеем

$$\frac{dS}{d\tau} = \left(\frac{dS}{d\tau}\right)^{F.P.} + \left(\frac{dS}{d\tau}\right)^{Ch} + \left(\frac{dS}{d\tau}\right)^{\delta}, \quad (10)$$

где вклад в изменение энтропии от оператора Фоккера – Планка имеет вид

$$\left(\frac{dS}{d\tau}\right)^{F.P.} = -k \int_0^{\infty} \left(\ln \frac{F(x,\tau)}{\sqrt{x}}\right) \frac{\partial}{\partial x}\left[\left(x - \frac{1}{2}\right) F + x \frac{\partial F}{\partial x}\right] dx = \quad (11)$$

$$= k \frac{\eta \tau_R}{1 + K\tau_R} L_1^{1/2}(x_0) \left[1 - e^{-(1+K\tau_R)\tau}\right] + k \frac{\sqrt{\pi}}{2} \eta \tau_R K \tau_R \sum_{m=1}^{\infty} \frac{\Gamma(m)}{\Gamma(m+3/2)} \left[L_m^{1/2}(x_0)\right]^2 \left(1 - e^{-(m+K\tau_R)\tau}\right)^2 +$$

$$+ 0\left[(K\tau_R)^2\right]$$

Как видно из (11) это изменение энтропии состоит из двух частей: первой - обратимое изменение энтропии за счет отвода (при $x_0 > x$) энергии в термостат, второе - необратимое производства энтропии.

$$\sigma(\tau) = k \frac{\sqrt{\pi}}{2} \eta \tau_R K \tau_R \sum_{m=1}^{\infty} \frac{\Gamma(m)}{\Gamma(m+3/2)} \left[L_m^{1/2}(x_0)\right]^2 \left(1 - e^{-(m+K\tau_R)\tau}\right)^2 \geq 0 \quad (12)$$

В стационарном состояние оператор Фоккера – Планка обеспечивает постоянный поток энтропии в термостат (при $x_0 > x$) и постоянно производит энтропию. Как видно из (12) производство энтропии стационарном состояния максимально и равно

$$\sigma(\tau) = k \frac{\sqrt{\pi}}{2} \eta \tau_R K \tau_R \sum_{m=1}^{\infty} \frac{\Gamma(m)}{\Gamma(m+3/2)} \left[L_m^{1/2}(x_0)\right]^2 \quad (13)$$

Таким образом, при релаксации из равновесного состояния в неравновесное стационарное состояние, производство энтропии, в соответствие с принципом Циглера, максимально.

3. Производство негэнтропии.

Как известно [1,2], в изолированной (замкнутой) системе $\sigma \geq 0$. Это соотношение является выражением второго закона в изолированной системе. Для открытой системы имеется выражение $dS/d\tau = d_iS/d\tau + d_eS/d\tau$, где $d_iS/d\tau = \sigma$, $d_eS/d\tau$ – обратимое изменение энтропии [2]. В открытой системе $\sigma > 0$, а $d_eS/d\tau$ может быть, как положительным, так и отрицательным. Однако, в работе [8], при моделировании колебательных процессов, при малых временах, обнаружено отрицательное производство энтропии. Хотя это мгновенное значение производства энтропии и временные средние не отрицательны. По-видимому, отрицательное производства энтропии — это выход в область нелинейной термодинамики. Это означает, что в нелинейной термодинамике необратимых процессов основное положение линейной термодинамике о том, что производство энтропии в необратимых процессах всегда положительно не применимо. По-видимому, в нелинейной термодинамике необратимые процессы не только производят энтропию, но могут и поглощать её, т.е. производить негэнтропию. Термин "производство негэнтропии" вводиться для того чтобы подчеркнуть поглощение энтропии необратимыми процессами.

Изолированная система релаксирует из неравновесного состояния в равновесное, при этом энтропия возрастает и производство энтропии положительно и равно нулю в равновесии. Открытая система релаксирует в стационарное неравновесное (динамически равновесное) состояние. В этом случае энтропия конечного состояния может быть, как больше, так и меньше энтропии начального состояния. Поэтому при релаксации из начального в конечное состояние энтропия будет либо производиться, либо поглощаться. Баланс энтропии в стационарном состоянии для исследуемой системы, описываемой уравнением (8) с граничными условиями (2), был показан в [9]. Здесь необходимо проанализировать баланс обратимой (потоковой) и необратимой (производимой/поглощаемой) энтропии в нестационарных условиях. Вклад в изменение энтропии от Фоккер - Планковского слагаемого получен выше в (11). Далее вычислим нестационарные вклады в изменения энтропии от химической реакции и $\delta$ - источника.

Изменение энтропии за счет химической реакции можно получить в виде

$$\left(\frac{dS}{d\tau}\right)^{Ch} = kK\tau_R \int_0^\infty \left(\ln\frac{F(x,\tau)}{\sqrt{x}}\right) F(x,\tau) dx = -\frac{3}{2}k\eta\tau_R + k\frac{\eta\tau_R K\tau_R}{1+K\tau_R} L_1^{1/2}(x_0)\left[1-e^{-(1+K\tau_R)\tau}\right] +$$

$$+ 0\left[(K\tau_R)^2\right] \qquad (14)$$

Химическая реакция, отводящая из системы частиц с текущей ФР, вносит в изменение энтропии только обратимые вклады, за счет уменьшения частиц и изменения энергии.

Наибольший интерес вызывает изменение энтропии за счет $\delta$ – источника, так как этот источник вносит в систему упорядоченные в пространстве энергий частицы

$$\left(\frac{dS}{d\tau}\right)^\delta = -k\eta\tau_R \int_0^\infty \left(\ln\frac{F(x,\tau)}{\sqrt{x}}\right)\delta(x-x_0)dx =$$

$$= \frac{3}{2}k\eta\tau_R - k\eta\tau_R L_1^{1/2}(x_0) - k\frac{\sqrt{\pi}}{2}\eta\tau_R K\tau_R \sum_{m=1}^\infty \frac{\Gamma(m)}{\Gamma(m+3/2)}\left[L_m^{1/2}(x_0)\right]^2 \left[1-e^{-(m+K\tau_R)\tau}\right] +$$

$$+ 0\left[(K\tau_R)^2\right] \qquad (15)$$

Первое слагаемое в (15) это возрастание энтропии за счет увеличения числа частиц, оно компенсируется отводом частиц химической реакцией. Второе слагаемое – поток энтропии, вызванной энергосодержанием вводимых частиц, компенсируется в стационарном состоянии суммой потоковых слагаемых в химической реакции и в Фоккер – Планковском операторе. Третье слагаемое является производством негэнтропии или необратимым поглощением энтропии, стимулированное в системе δ – источником и имеет вид

$$\tilde{\sigma}(\tau) = -k\frac{\sqrt{\pi}}{2}\eta\tau_R K\tau_R \sum_{m=1}^{\infty}\frac{\Gamma(m)}{\Gamma(m+3/2)}\left[L_m^{1/2}(x_0)\right]^2\left[1-e^{-(m+K\tau_R)\tau}\right] \leq 0 \qquad (16)$$

Под производство негэнтропии следует понимать отрицательное производство (поглощение) положительной энтропии, а не наоборот. Оно компенсирует в стационарном неравновесном состоянии производство энтропии. Таким образом, общепринятое мнение, что производимая в стационарном неравновесном (динамически равновесном) состояние энтропия компенсируется оттоком энтропии (негэнтропией), в рассмотренной модели не выполняется.

4. Второй закон в открытой системе.

Как было отмечено выше, в изолированной системе второй закон термодинамики определяется неравенством σ≥0. "Несмотря на чрезвычайно важную информацию, содержащуюся в этом неравенстве, она недостаточна по двум причинам. Во-первых, мало только знать, что энтропия возрастает. Надо установить источники возникновения энтропии и их мощность, то есть найти скорость возрастания энтропии (величину σ). Во-вторых, общая формулировка второго начала не говорит, что же происходит в неизолированных (открытых) системах, способных обмениваться веществом и энергией с окружающей средой" [1]. В предыдущих разделах вычислены вклады всех источников в изменение энтропии в релеевском газе с источниками (открытая система), т.е. выполнена первая часть, сформулированной в цитате из [1], задачи. Теперь, чтобы выполнить вторую часть задачи, необходимо вычислить результат совместного действия всех источников энтропии и из этого выражения получить вариант второго закона в открытой системе.

Таким образом, чтобы сформулировать второй закон в исследуемой (открытой) системе используем явные нестационарные вклады в изменения энтропии каждого слагаемого правой части уравнения (8), полученные выше. Подставим в (10) выражения (11), (14) и (15). Тогда полное изменение энтропии при релаксации из равновесного состояния в неравновесное стационарное состояние можно записать как

$$\frac{dS}{d\tau} = -k\eta\tau_R L_1^{1/2}(x_0)e^{-(1+K\tau_R)\tau} -$$

$$-k\frac{\sqrt{\pi}}{2}\eta\tau_R K\tau_R \sum_{m=1}^{\infty}\frac{\Gamma(m)}{\Gamma(m+3/2)}\left[L_m^{1/2}(x_0)\right]^2\left[1-e^{-(m+K\tau_R)\tau}\right]e^{-(m+K\tau_R)\tau} \qquad (17)$$

Первое слагаемое в (17) описывает обратимое изменение энтропии, второе – необратимое изменение энтропии. Это слагаемое в (17) отрицательно, это связано с тем, что выбранная модель релаксирует из равновесного (с максимальной энтропией) в стационарное неравновесное (с меньшей энтропией) состояние. Это слагаемое можно было бы назвать обобщенным производством энтропии, имея в виду, что оно является суммой производств энтропии и негэнтропии, т.е.

$$\Xi(\tau)=\sigma(\tau)+\tilde{\sigma}(\tau).$$

Тогда на основе анализа баланса производств энтропии и негэнтропии можно сформулировать один из возможных вариантов второго закона для открытых систем следующим образом [10], *при релаксации открытой системы в неравновесное стационарное состояние, обобщенное производство энтропии убывает по абсолютному значению и равно нулю в стационарном неравновесном состояние.*

$$|\Xi| \geq 0$$

Знак обобщенного производства энтропии определяет направление релаксации, если оно положительно система релаксирует в направления равновесного состояния, если отрицательно – система отдаляется от равновесного состояния.

## 5. Заключение.

Таким образом, в открытой неравновесной системе при релаксации в более упорядоченное состояние производится больше негэнтропии, чем энтропии, и выполняется ПМакПЭ, а при релаксации в более неупорядоченное состояние производится больше энтропии и выполняется ПМинПЭ. В обоих случаях в стационарном состояние достигается баланс производств энтропии и негэнтропии. Если в изолированной системе, при релаксации к равновесию, производство энтропии убывает и равно нулю в равновесии, то в открытой системе, при релаксации к стационарному состоянию, модуль обобщенного производства энтропии убывает и равен нулю в стационарном состояние.

## Литература.